\documentclass[%
reprint,
 amsmath,amssymb,
 prl,
]{revtex4-1}

\usepackage{graphicx}% Include figure files
\usepackage{dcolumn}% Align table columns on decimal point
\usepackage{bm}% bold math
\usepackage{natbib}
\usepackage{soul}
\usepackage{color}
\usepackage{hyperref}% add hypertext capabilities
\begin{document}

\title{Mesoscopic wave physics in fish shoals}% Force line breaks with \\

\author
{Benoit Tallon}
\affiliation{Univ. Grenoble Alpes, CNRS, ISTerre, 38000 Grenoble, France}

\author
{Philippe Roux}
\email[]{To whom correspondence should be addressed; E-mail: philippe.roux@univ-grenoble-alpes.fr}
\affiliation{Univ. Grenoble Alpes, CNRS, ISTerre, 38000 Grenoble, France}

\author
{Guillaume Matte}
\affiliation{iXblue, Sonar division, 13600 la Ciotat, France}

\author
{Jean Guillard}
\affiliation{Univ. Savoie Mont Blanc, INRA, CARRTEL, 74200 Thonon-les-Bains, France}

\author
{Sergey E. Skipetrov}
\affiliation{Univ. Grenoble Alpes, CNRS, LPMMC, 38000 Grenoble, France}

%% \author
%% {Benoit Tallon,$^{1}$ Philippe Roux,$^{1\ast}$ Guillaume Matte,$^{2}$ Jean Guillard,$^{3}$\\ and Sergey E. Skipetrov$^{4}$\\
%% \normalsize{$^{1}$Univ. Grenoble Alpes, CNRS, ISTerre, 38000 Grenoble, France}\\
%% \normalsize{$^{2}$iXblue, Sonar division, 13600 la Ciotat, France}\\
%% \normalsize{$^{3}$Univ. Savoie Mont Blanc, INRA, CARRTEL, 74200 Thonon-les-Bains, France}\\
%%\normalsize{$^{4}$Univ. Grenoble Alpes, CNRS, LPMMC, 38000 Grenoble, France}\\
%%\normalsize{$^\ast$To whom correspondence should be addressed; E-mail: philippe.roux@univ-grenoble-alpes.fr.}\\
%%}

\date{\today}% It is always \today, today,

\begin{abstract}

Ultrasound scattered by a dense shoal of fish undergoes mesoscopic interference, as is typical of low-temperature electrical transport in metals or light scattering in colloidal suspensions.
Through large-scale measurements in open sea, we show a set of striking deviations from classical wave diffusion making fish shoals good candidates to study mesoscopic wave phenomena. The very good agreement with theories enlightens the role of fish structure on such a strong scattering regime that features slow energy transport and brings acoustic waves close to the Anderson localization transition.

\end{abstract}

                              %PACS numbers: 43.35.+d, 43.20.+g, 62.30.+d
%Use showkeys class option if keyword
\maketitle

%\tableofcontents

Since the late 1980's, physicists have achieved great progress in the fabrication of strongly disordered materials that would allow for Anderson localization of `classical' waves (e.g., light, microwaves, sound) in three dimensions \cite{lagendijk09,wiersma13,skip16}. Anderson localization is a halt of propagation due to disorder \cite{anderson58,abrahams10}. Although very few experiments have succeeded \cite{hu08,cobus16,cobus18}, these studies have revealed the `mesoscopic' interference phenomena that are analogous to that of low-temperature electrical transport \cite{vanrossum99,akkermans07}: weak localization \cite{albada85,wolf85}, universal conductance fluctuations \cite{feng91,scheffold98}, strong fluctuations and long-range correlations of scattered intensity \cite{hilder14,star18}. These studies also led to the discovery of new phenomena, such as the slowing down of transport due to scattering resonances \cite{albada91}, random lasing \cite{wiersma08}, mean path length invariance \cite{savo17}, and transverse localization of transmission eigenchannels \cite{yilmaz19}.Such a set of mesoscopic phenomena have never been observed apart from laboratory experiments, and even less with living matter, as they require carefully designed disordered samples. In acoustics, three-dimensional mesoscopic phenomena have been observed exclusively in so-called `mesoglasses' \cite{hu08}. Many studies have considered coated particles suspended in a host matrix in both optics and acoustics \cite{liu2000,naraghi2015} because of their interesting scattering properties. However arduous synthesis and weak stability make those particles rarely employed. Can natural complex media be inspiring for the design of such model systems? Do they scatter waves strongly enough to observe non diffusive wave transport?

%% Vitesse de l'energie la plus faible
%% Objets coqués => on peut tirer avantage des différents consitutants
%% Est-ce que on peut toruver des analogues à ces systemes modèles dans la nature?

Here, we show that shoals of fish trapped in large cages---an example of live, active matter---allow the examination of various mesoscopic interference phenomena in ultrasound scattering for fish densities that are comparable to those encountered in natural fish schools at sea. Fish swim bladder (an organ which allows fish to control their buoyancy) is analogous to an air bubble and thus strongly scatters ultrasonic waves. This strong scattering has been useful for several decades for fish counting with ultrasounds in the single scattering regime \cite{simmonds08}.
The present study focuses on dense shoals in which single scattering assumption is irrelevant.
Comparison with multiple scattering theories reveals the impact of the complex fish structure that can be seen as a coated air bubble.
For different fish densities, the scattering strength of fish shoals is demonstrated \textit{via} measurements of long-range correlations or non-Rayleigh distribution of the intensity speckle, as well as \textit{via} the dynamic coherent backscattering effect, revealing the lowest energy velocity ever observed in underwater acoustics.
Because of their quasi-random movement, fish are also interesting for configurational averaging where spatial ergodicity is usually assumed for laboratory experiments.

\begin{figure*}[t]
%%\vspace*{-0.5cm}
%% \hspace*{-0.55cm}
\includegraphics[width=0.8\textwidth]{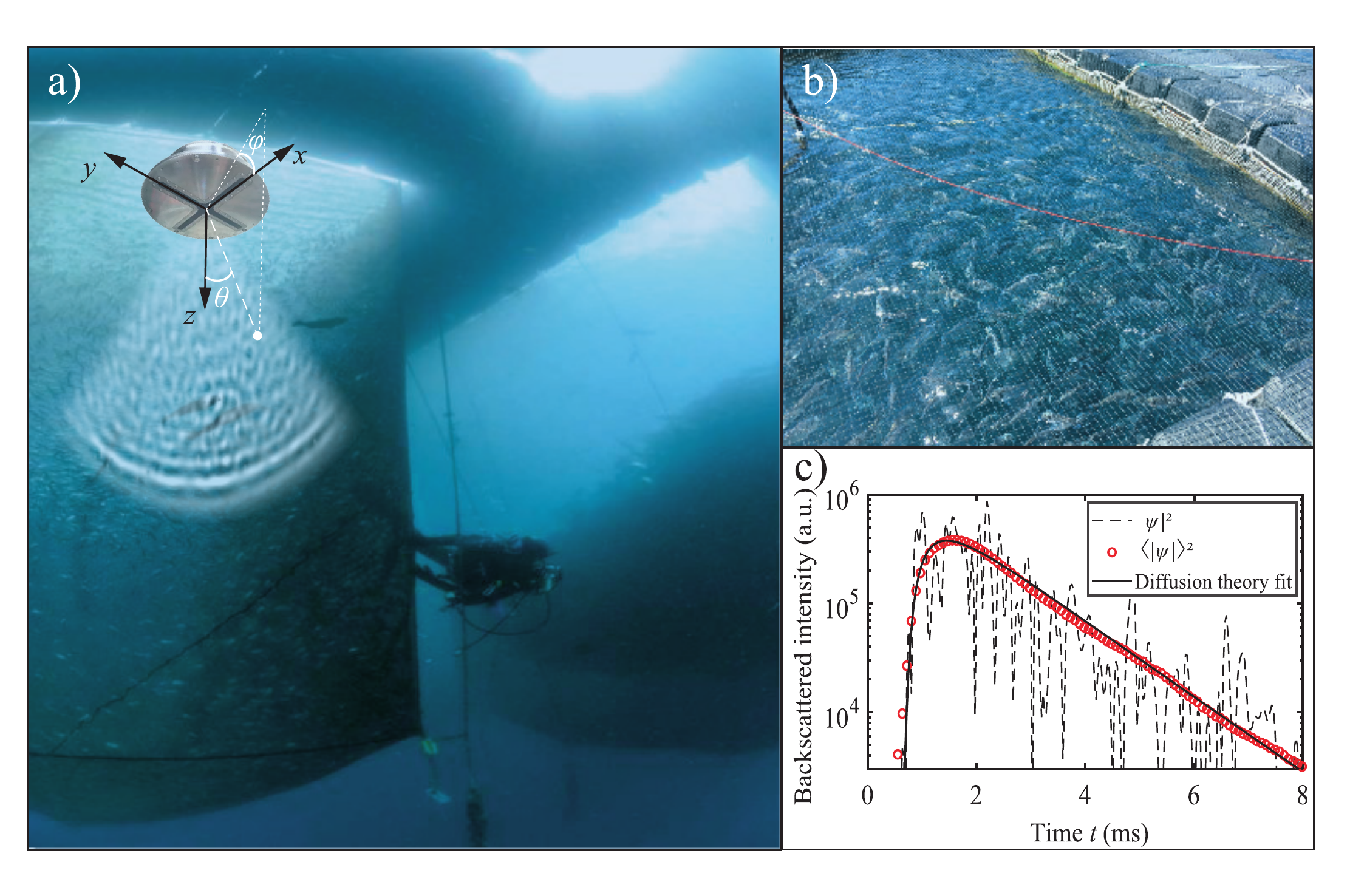}
%%\vspace*{-1.3cm}
\caption{\label{fig_scheme}
(a, b) Photographs taken underwater (a) and at the surface (b) of an open-sea cage used in these experiments (\textit{courtesy of S. Pasta}). (c) Intensity (dashed line) and mean intensity (symbols) of the backscattered acoustic signal after emission of a short pulse. Solid line, diffusion theory fit.
}
\end{figure*}

To ensure fish shoal control and to prevent avoidance reactions of the fish at sea \cite{makris09}, we perform acoustic measurements in large ($\sim 5$ m $\times 5$ m $\times 5$ m) open-sea fish cages that are typical of fish farms (Fig.\ \ref{fig_scheme}a, b). A cage typically contains several tens of thousands of fish at a mean density of $10$--$100$ fish per m$^3$. The individual fish mass ranges from $10$ g for fish larvae up to $1$ kg for mature fish.
The corresponding fish volume fraction $\phi$ ranges from $1\%$ to $10\%$. Much higher $\phi$, up to $30$\%, can be reached with fish farms that practice intensive fish farming.

We measure the reflection of short acoustic pulses ($\sim 0.1$ ms) that are emitted by a cross-shaped array of $2 \times 64$ acoustic transducers, as shown in Figure\ \ref{fig_scheme}a, at a central frequency of $f = 150$ kHz.
This is far from the swim-bladder resonance \cite{sm}. Successive measurements are repeated several thousands of times at a rate of $\sim 30$ shots per second. The natural fish motion at a speed of the order of $5$ cm/s is sufficiently slow to ensure that the fish can be considered as immobile during each single shot. At the same time, the fish motion produces independent fish configurations over time, effectively providing us with a huge number of statistically independent measurements that correspond to different configurations of the fish in the cage.

For a single incident pulse, the backscattered acoustic pressure field $\psi(\mathbf{r},t)$ and the intensity $I(\mathbf{r},t) = |\psi(\mathbf{r},t)|^2$ measured by a transducer at a position $\mathbf{r}$  fluctuate widely (see Fig.\ \ref{fig_scheme}c, dashed line, for typical experimental data), whereas the mean over many shots (and hence many different fish configurations) and over all of the transducers yields a smooth \textit{coda}, as shown by the symbols in Figure\ \ref{fig_scheme}c. This is a direct signature of the multiple scattering, and it can be understood by viewing wave transport as a random walk with a velocity $v$, a step length between scattering events $\ell$ (the scattering length, or mean free path), and an isotropization distance $\ell^* \geq \ell$ (the transport mean free path). For propagation distances greater than $\ell^*$, the ensemble mean intensity $\langle| \psi(\mathbf{r},t)|^2 \rangle$ obeys a diffusion equation with diffusivity $D = v\ell^*/3$ \cite{akkermans07}. This equation provides an excellent description of $\langle| \psi(\mathbf{r},t)|^2 \rangle$ in Figure\ \ref{fig_scheme}c (solid line).
This diffuse behavior of $\langle I(\mathbf{r},t) \rangle$ is accompanied by circular Gaussian statistics of $\psi(\mathbf{r},t)$.

\begin{figure*}[t]
%%\vspace*{-0.5cm}
%% \hspace*{-0.55cm}
\includegraphics[width=0.8\textwidth]{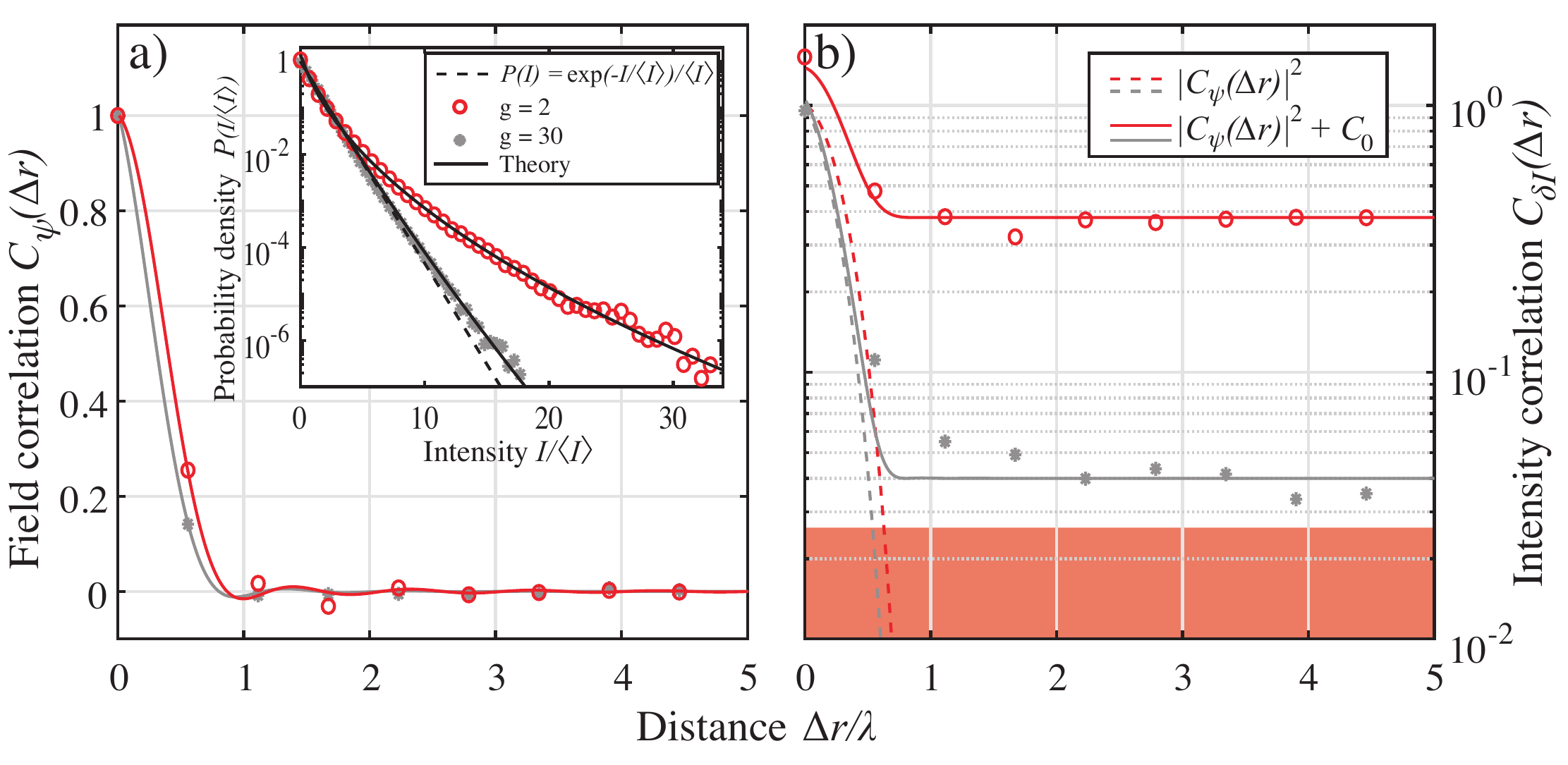}
%%\vspace*{-1.3cm}
\caption{\label{fig_stat}
Spatial correlation of field $\psi$ (a) and intensity fluctuations $\delta I$ (b) for weakly (gray asterisks) and strongly (red circles) scattering fish shoals. The lines show the theoretical fits to the data. While $C_{\psi}(\Delta r)$ rapidly decays to zero for both shoals, $C_{\delta I}(\Delta r)$ is long-lasting and remains appreciable even for $\Delta r = 5 \lambda$, especially for the stronger-scattering shoal. The colored area in (b) represents the noise level.
(a) inset: the probability density of the normalized intensity $I/\langle I \rangle$ (symbols) that deviates from the Rayleigh law $P(I) = \exp(-I/\langle I \rangle)/\langle I \rangle$ expected for weak scattering (dashed line). Solid lines are theoretical fits \cite{vanrossum99} with the dimensionless conductance $g$ as a free-fit parameter.
}
\end{figure*}

Mesoscopic effects manifest as deviations from the diffusion picture of propagation, due to interference of waves scattered along different paths inside the disordered fish aggregation \cite{akkermans07,vanrossum99}. We first analyze the statistics of the time-integrated (stationary) wave field $\psi(\mathbf{r})$ and intensity $I(\mathbf{r}) = |\psi(\mathbf{r})|^2$. Correlation functions $C_{\psi}(\Delta r) = \langle \psi(\mathbf{r}) \psi(\mathbf{r}+\Delta \mathbf{r})^* \rangle/\langle |\psi(\mathbf{r})|^2 \rangle$ and $C_{\delta I}(\Delta r) = \langle \delta I(\mathbf{r}) \delta I(\mathbf{r}+\Delta \mathbf{r}) \rangle/\langle I(\mathbf{r}) \rangle^2$ [where $\delta I(\mathbf{r}) = I(\mathbf{r}) - \langle I(\mathbf{r}) \rangle$] of the field and intensity fluctuations, respectively, are shown in Figure\ \ref{fig_stat} for two representative shoals that feature weak (gray asterisks) and strong (red circles) scattering. Weak scattering occurs for the fish fry (i.e., sea bream with mean weight $W = 10$ g and shoal density $\eta \sim 6$ kg/m$^3$), whereas strong scattering occurs for the dense shoal of adult sea bream ($W = 284$ g; $\eta \sim 23$ kg/m$^3$).
$C_{\psi}(\Delta r)$ is short-range for both shoals and can be reasonably well fitted according to theory that takes into account the finite size of our acoustic transducers \cite{sm} and yields the scattering lengths $\ell \sim \lambda \simeq 1$ cm as the best-fit parameters.
These small $\ell$ suggest that the Anderson localization of sound expected for $2\pi\ell/\lambda \lesssim 1$ (the Ioffe-Regel criterion) \cite{anderson58,abrahams10,lagendijk09} would be reachable in denser fish shoals.
In contrast to $C_{\psi}(\Delta r)$, the intensity correlation function $C_{\delta I}(\Delta r)$ features a long-range component $C_0$ that does not vanish even for $\Delta r \gg \lambda$, $\ell$. In our notation, $C_0$ incorporates contributions from wave interference in the bulk [denoted as $C_2$ and $C_3$ in the literature \cite{vanrossum99,akkermans07}] as well as the near transducers (the genuine $C_0$ \cite{shapiro99}) because our experiments do not allow these to be distinguished.
$C_0 \simeq 0.4$ for the dense fish shoal indicates a breakdown of wave diffusion where $C_{\delta I}(\Delta r) = |C_{\psi}(\Delta r)|^2$ would be expected. This breakdown is also
confirmed by an analysis of the intensity probability density function, as shown in Figure\ \ref{fig_stat}a, inset. Fitting of the data to theory \cite{vanrossum99} allow an effective dimensional conductance $g$ to be attributed to each fish shoal \cite{sm}. The variance of the intensity fluctuations is given by $\langle \delta I(\mathbf{r})^2 \rangle/\langle I(\mathbf{r}) \rangle^2 = 1 + 4/3g$ \cite{vanrossum99}. Thus, $g$ is a measure of deviations of the scattered field $\psi(\mathbf{r})$ from the Gaussian statistics, which implies $\langle \delta I(\mathbf{r})^2 \rangle/\langle I(\mathbf{r}) \rangle^2 = 1$. The large $C_0 \simeq 0.4$ and small $g \simeq 2$ for the high-density fish shoal signal substantial deviations from the diffusion picture of wave propagation and confirm the proximity of the Anderson localization regime, for which $g < 1$ is expected \cite{chabanov00}.

\begin{figure*}[t!]
%%\vspace*{-0.5cm}
%% \hspace*{-0.55cm}
\includegraphics[width=0.8\textwidth]{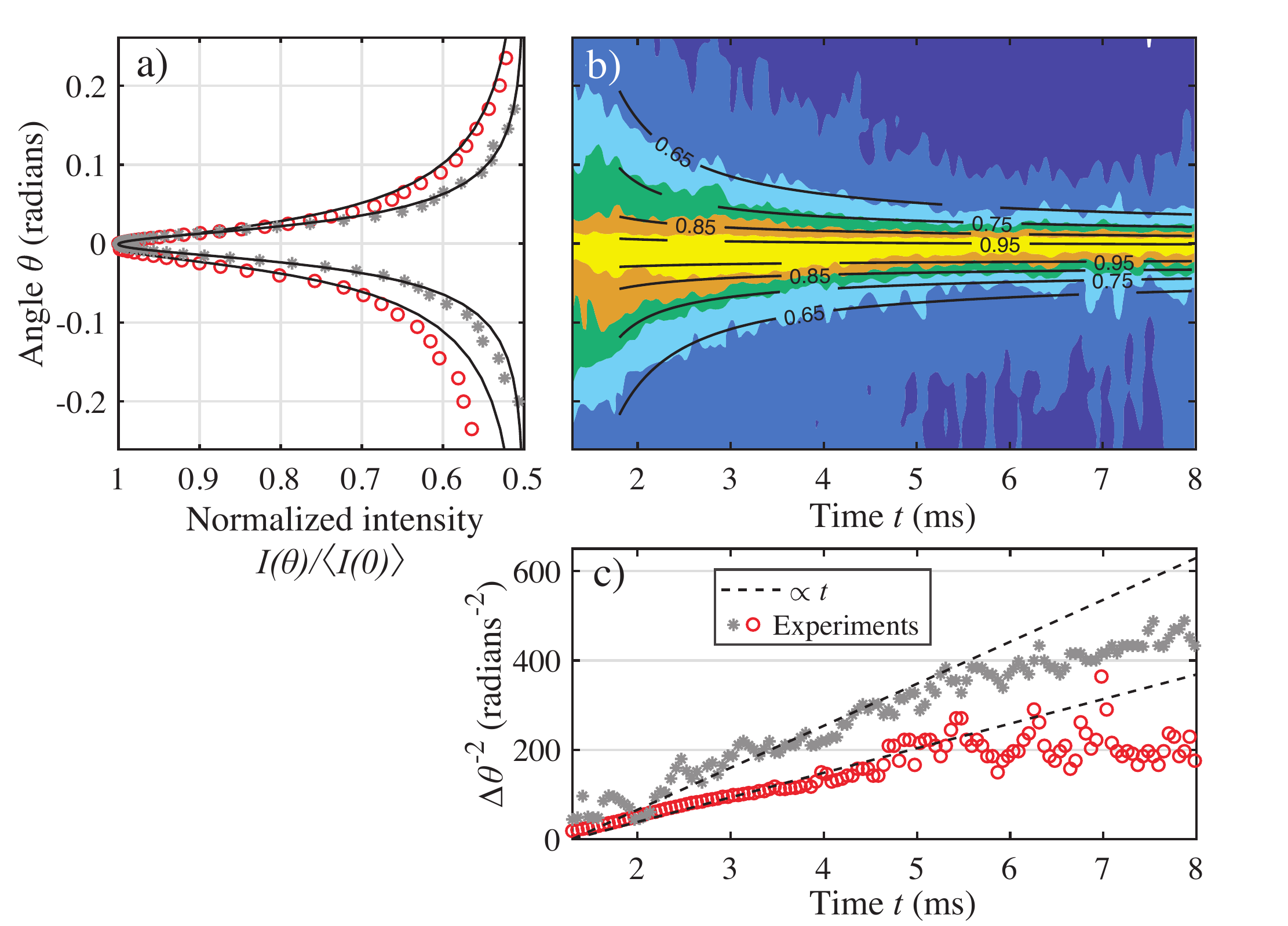}
%%\vspace*{-1.3cm}
\caption{\label{fig_cbs}
(a) Stationary coherent backscattering (CBS) profiles for weakly (gray asterisks) and strongly (red circles) scattering fish shoals. The lines are the theoretical fits. (b) Dynamic CBS profile for the strongly scattering shoal. (c) Time evolution of the cone width $\Delta \theta(t)$ of the dynamic CBS peaks. The linear increase in $\Delta \theta^{-2}$ that is expected from diffusion theory is shown by the dashed lines.
}
\end{figure*}

Coherent backscattering (CBS) represents a mesoscopic effect \textit{par excellence} that has been measured for light \cite{albada85,wolf85}, ultrasound \cite{tourin97}, matter \cite{jendr12} and seismic \cite{larose04} waves. This is due to constructive interference of waves following time-reversed pairs of paths, and manifests in the static regime as the doubling of the mean scattered intensity in a narrow angular range $\Delta \theta \sim \lambda/\ell^*$ around the back-scattering direction \cite{akker86}.
Examples of our CBS measurements are shown in Figure\ \ref{fig_cbs}a, for two cages that contain either adult sea bream at a low density ($W = 320$ g; $\eta \sim 15$ kg/m$^3$) or a dense shoal of croaker fish ($W = 886$ g; $\eta \sim 24$ kg/m$^3$). The theoretical fits to the data (see Supplementary text) provide the best-fit values of $\ell^* = 1.7$ cm and $\ell^* = 0.7$ cm for the lower and higher fish densities, respectively.
These small values of $\ell^*$ support our conclusion on the strong multiple scattering of ultrasound in the fish shoals considered.

The diffusivity $D$ can be estimated from the \textit{dynamic} CBS profile, as shown in Figure\ \ref{fig_cbs}b. The CBS cone width $\Delta\theta(t)$ follows the theoretical behavior $\Delta \theta^{-2} \propto D t$ expected for wave diffusion \cite{akkermans07,tourin97} up to $t \simeq 5$ ms (Fig.\ \ref{fig_cbs}c). The linear growth of $\Delta \theta^{-2}$ with time slows at longer times, again indicating strong mesoscopic interference effects and the closeness to the Anderson localization transition \cite{cobus16}.

Combining $D$ estimated above with $\ell^*$ from the static CBS, we obtain a surprisingly low energy transport velocity $v = 3D/\ell^* \simeq 35$ m/s. This value is much lower than the speed of sound in either water ($v_0 \simeq 1500$ m/s) or air
($340$ m/s)---the two values that might serve as a reference for scattering by an air-filled swim bladder in (possibly) bubbly water.
While it has been demonstrated that resonant scattering can slow down diffusive wave transport \cite{albada91,schriemer97,tallon17}, narrow-band resonance effects are not clear here since low $v$ values are obtained for any fish size.
The only possibility to explain this is to consider the solid multi-layer structure of a fish (see supplementary text) and, in particular, to invoke the slow speed $v_s \sim 10$ m/s of the shear waves in the fish flesh.
By assuming equipartition of the scattered intensity between longitudinal waves in water and shear waves in fish \cite{wearver82}, we consider that the wave speed is $v_s$ in the fish body and $v_0$ in between two fish. Then, averaging this along a path that traverses the fish shoal yields
$v = [1 + \phi^{1/3}/(1-\phi)^{1/3}]/[1/v_0 + \phi^{1/3}/(1-\phi)^{1/3}/v_s] \simeq 30$ m/s for the fish volume fraction $\phi \simeq 10$\%, which is in agreement with independent measurements provided by the sea-farm manager.
We emphasize that this dramatic slowing down of the ultrasound is not related to the scattering resonances of the fish, and thus cannot be explained by known, resonant mechanisms \cite{albada91,cowan11}.

From an applied standpoint for aquaculture, determination of $v$ via dynamic CBS measurements allows the estimation of the shoal density through the fish volume fraction $\phi$. Similarly, measurements of $C_{\psi}(\Delta r)$ and the stationary CBS yield $\ell$ and $\ell^*$, which are related to the scattering and transport cross-sections $\sigma$ and $\sigma^*$ of an individual fish, from which a mean fish length can be estimated. The knowledge of both the mean fish length and the fish shoal density opens new perspectives for noninvasive biomass estimation of dense fish shoals.

In conclusion, ultrasound scattering in fish shoals
under conditions close to those encountered in nature   
show such mesoscopic wave phenomena as long-range correlations of scattered wave intensity, CBS, and the slowing down of the diffusion.
These phenomena indicate that transition to the Anderson localization might be within reach in experiments with denser fish shoals.
The extremely slow energy transport velocity emphasizes the importance of the fish solid structure in the multiple scattering of ultrasound---a phenomenon that has been overlooked up to now.
Furthermore, the alliance of mesoscopic wave physics and fisheries acoustics has the potential of being used for monitoring fish biomass, which at present is restricted to the single scattering regime.
When transposed to the open sea, CBS measurements might also be applied to study strong fish density variations during day-night schooling transitions \cite{makris09}.

The authors thank Prof. J. H. Page for valuable discussions about transport intensity velocity.

\end{document}

\title{Supplemental material for \\ ``Mesoscopic wave physics in fish shoals"}% Force line breaks with \\

\author
{Benoit Tallon,$^{1}$ Philippe Roux,$^{1\ast}$ Guillaume Matte,$^{2}$ Jean Guillard,$^{3}$\\ and Sergey E. Skipetrov$^{4}$\\
\normalsize{$^{1}$Univ. Grenoble Alpes, CNRS, ISTerre, 38000 Grenoble, France}\\
\normalsize{$^{2}$iXblue, Sonar division, 13600 la Ciotat, France}\\
\normalsize{$^{3}$Univ. Savoie Mont Blanc, INRA, CARRTEL, 74200 Thonon-les-Bains, France}\\
\normalsize{$^{4}$Univ. Grenoble Alpes, CNRS, LPMMC, 38000 Grenoble, France}\\
\normalsize{$^\ast$To whom correspondence should be addressed; E-mail: philippe.roux@univ-grenoble-alpes.fr.}\\
}

\date{\today}% It is always \today, today,

                              %PACS numbers: 43.35.+d, 43.20.+g, 62.30.+d
%Use showkeys class option if keyword
\maketitle
%\tableofcontents
\section{Experimental set-up}

\subsection{Open-sea cages}

The sea cages are in the Mediterranean Sea (Cannes, France). The distance to the sea bottom is $z =$ 6.5 m for the smallest cage and  $z =$ 15 m for the biggest cage, from their lowest points.
The parameters relating to the shoals and cages (as provided by the farm manager) are detailed in Table \ref{tab1}. To maintain the organic label of the farm and to avoid the need for drug treatments, the fish densities in these cages are lower than in intensive farming facilities (where mass densities $\eta$ can reach 100 kg/m$^3$).

\renewcommand{\arraystretch}{1.2}
\begin{table}
\begin{center}
\begin{tabular}{|l|c|c|c|c|}
\hline
\cellcolor[HTML]{9B9B9B}&$N$&$W$ (g)&$\eta$ (kg/m$^3$)&$V$ (m$^3$)\\ \hline
\begin{tabular}[c]{@{}l@{}}C1 (sea bream, adults, Fig. 1c)\end{tabular} &5,000& 500 & 7& $343$ \\ \hline
\begin{tabular}[c]{@{}l@{}}C2 (sea bream, fry, Fig. 2)\end{tabular} &75,000& 10 & 6& $125$ \\ \hline
\begin{tabular}[c]{@{}l@{}}C3 (sea bream, adults, Fig. 2)\end{tabular} & 10,080 & 284 & 23 & $125$ \\ \hline
\begin{tabular}[c]{@{}l@{}}C4 (sea bream, adults, Fig. 3)\end{tabular} & 6,000 & 320 & 15 & $125$ \\ \hline
\begin{tabular}[c]{@{}l@{}}C5 (croakers, adults, Fig. 3)\end{tabular} & 13,900 & 886 & 24 & $512$ \\ \hline
\end{tabular}
\end{center}
\caption{\label{tab1} Number of fish $N$ and mean fish weight $W$ and shoal density $\eta$ for the five open sea cages considered in this study. $V$, cage volume.}
\end{table}
\renewcommand{\arraystretch}{1}

The fish fry are raised in in-shore tanks and transferred to the open-sea cages when their weight is about 5 g.
The feeding procedures are controlled to obtain a calibrated fish size.
Hormones are not used in this farm, so 18 months to 3 years are required to grow the fish to their adult size (which corresponds to the standard commercial weight for a fish fillet).
Species choice is also an organic label constraint as sea bream, croakers and sea bass (not studied here) are native species of the Cannes region.

\subsection{Ultrasonic antenna}

The acoustic device used here is a Mills cross-shaped antenna that consists of two perpendicular linear arrays, each with 64 ultrasonic piezoelectric transducers.
Each transducer is narrow band (bandwidth, 10 kHz) with a central frequency of 150 kHz, which corresponds to a wavelength $\lambda \simeq$ of 1 cm in water. The transducers are square in shape and have sides of 0.5 cm $\simeq \lambda$/2.
The antenna (expressly designed by the iXblue company for aquaculture monitoring) is placed just below the water surface, facing the sea bottom.

\begin{figure}[t]
%%\vspace*{-0.5cm}
%% \hspace*{-0.55cm}
\includegraphics[width=\columnwidth]{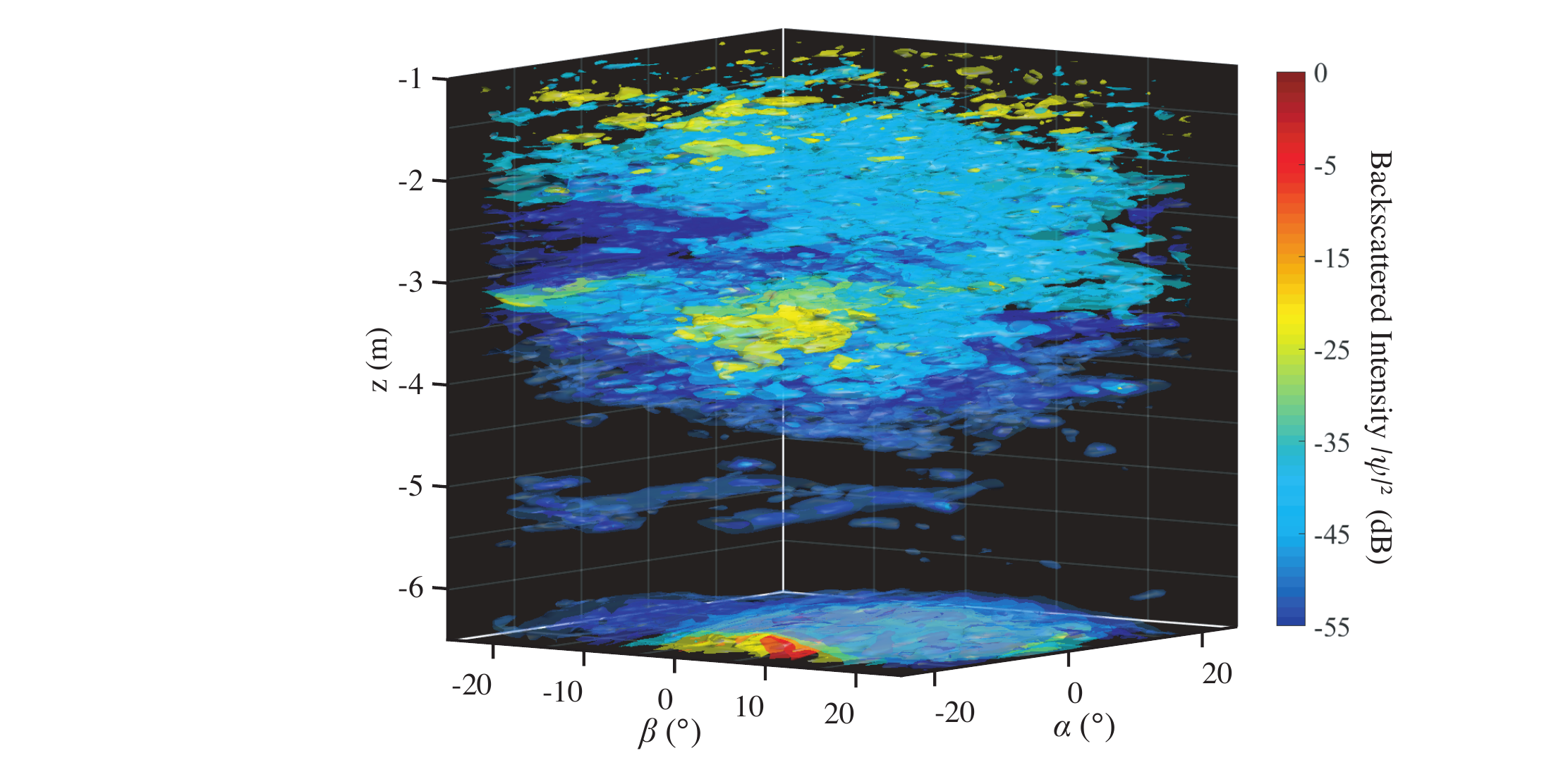}
%%\vspace*{-1.3cm}
\caption{\label{figS1}
Three-dimensional acoustic scan of an open sea cage (C2). The cage net is just visible for $z =$ 5 m, as well as the sea bottom for $z =$ 6.5 m.}
\end{figure}

The Mills cross array allows us to obtain a rapid three-dimensional representation of the ultrasound intensity back-scattered by the fish shoals, by emitting a series of tilted plane waves through one array (with controlled direction $\alpha$) and recording through the other array. The measurements are projected into the angular $\beta$ space in the post-processing. This acquisition sequence lasts less than 1.5 s and provides the time-evolving back-scattered acoustic intensity in the $\alpha$--$\beta$ angular space (Fig. \ref{figS1}).

Two types of measurement are used to characterize the shoal: (i) to measure spatial correlations (Fig. 2), only the central element of the antenna emits the acoustic wave, whereas all of the 128 elements are used to record the back-scattered signal; (ii) to measure the intensity profile (Fig. 1c), the probability density (Fig. 2a, inset) and the coherent back-scattering (Fig. 3), all of the 128 transducers of the antenna are used as both transmitters and receivers, without applying any emission delay ($\alpha = 0$).
For all of the measurements, the duration of the emitted pulse is about 0.1 ms and the back-scattered waves are recorded over 25 ms immediately after emission.
After each acquisition cycle, the system remains inactive for 10 ms to ensure that no residual acoustic signal is detected at the beginning of the next cycle.

Averaging is performed over 2,000 to 20,000 acquisitions (roughly 10-min-long acquisitions), to ensure that many different independent spatial configurations of the fish in the cages are probed.
Using an acoustic set-up with phase sensitive detectors allows us to directly access the wave field $\psi(\mathbf{r},t)$ and not only the intensity $I(\mathbf{r},t) = |\psi(\mathbf{r},t)|^2$, as in standard optical experiments.
The mean amplitude $\langle \psi(\mathbf{r},t)\rangle$ of the detected ultrasound is subtracted from each acquisition $\psi(\mathbf{r},t)$ to exclude any residual ballistic contribution (such as ringing artifacts or echos from the cage net).

\section{Data processing and fitting}

\subsection{Diffusion approximation for the average intensity}

The mean back-scattered intensity $\langle I(t) \rangle$ (Fig. 1c, symbols) is obtained after averaging $I(\mathbf{r},t)$ over the acquisitions and positions $\mathbf{r}$ of the transducers of the antenna.
The theory used to fit the experimental data is derived from a diffusion equation for $\langle I(\mathbf{r},t) \rangle$ \cite{akkermans07} that is solved for a disordered medium that occupies the half-space $z > 0$, with a delta-function source at $z = z'$ \cite{Carslaw1959}:
\begin{equation}\label{eqS1}
\langle I(t) \rangle = \frac{I_0}{2\pi}\int_{-\infty}^{+\infty}\frac{z_0\textrm{exp}(-\gamma_0 z')}{D(1+\gamma_0 z_0)}\textrm{exp}(-\textrm{i}\Omega t)d\Omega,
\end{equation}
where $\gamma_0^2(\Omega) = \frac{-\textrm{i}\Omega}{D} + \frac{1}{D\tau_a}$, $\tau_a$ is the characteristic absorption time, $z_0 = \frac{2}{3}\frac{1+R}{1-R}\ell^*$ is the extrapolation length, with $R =$ 0.99 as the reflection coefficient for the water/air interface, $D$ is the sound diffusion coefficient, and $\ell^*$ is the transport mean free path. $\ell^*$ and $D$ are deduced from the static and dynamic coherent back-scattering (CBS) measurements, respectively, while the fitting parameters are $z' =$ 1.8 mm, $\tau_a$ = 1.4 ms, and the overall amplitude is $I_0$ = 5 $\times$ 10$^6$.

\subsection{Spatial correlations and probability density}

The field correlation function $C_\psi(\Delta r) = \langle \psi(\mathbf{r}) \psi(\mathbf{r}+\Delta \mathbf{r})^* \rangle/\langle |\psi(\mathbf{r})|^2 \rangle$ is calculated in the frequency domain and averaged over the acquisitions and frequencies of the antenna bandwidth. The theoretical expression used to fit the data in Fig. 2a is:
\begin{equation}\label{eqS2}
\begin{split}
C_\psi(\Delta r) = \frac{\textrm{exp}(-\Delta r/2\ell)}{(k_0a)^2}
\{ k_0(a-\Delta r) \textrm{Si}\left [k_0(a-\Delta r)\right ] +  k_0(a+\Delta r) \textrm{Si}\left [k_0(a+\Delta r)\right ]\\
+ 
2 \left [\textrm{cos}(k_0a) - 1\right ]\textrm{cos}(k_0\Delta r) -
2k_0\Delta r \textrm{Si}(k_0\Delta r) \},
\end{split}
\end{equation}
where Si is the sine integral function, $k_0$ is the wave number in pure water, $a$ is the transducer size, and $\ell$ is the scattering length (mean free path). This expression is obtained by averaging the standard expression for field correlation in a disordered medium \cite{akkermans07},
\begin{equation}\label{eqS3}
C_\psi(\Delta r) = \frac{\sin k_0 \Delta r}{k_0 \Delta r} \exp\left(-\frac{\Delta r}{2\ell} \right),
\end{equation}
over transducers of size $a$ for $k_0 \ell \gg 1$, and assuming that transducers are one-dimensional (i.e., each transducer is a line of length $a$). We checked that relaxing this one-dimensional approximation and considering square transducers of size $a \times a$ yields data that are very close to those following from Equation (\ref{eqS2}) for the ranges of our parameters, albeit with a slightly different value of $a$. Using $a$ and $\ell$ as free-fit parameters, we obtain the fits of Figure 2a with $\ell =$ 1 cm and $a = 3\lambda/4$.  

The fit to the intensity correlation function $C_{\delta I}(\Delta r)$ in Figure 2b is obtained using
\begin{equation}\label{eqS4}
C_{\delta I}(\Delta r) = |C_\psi(\Delta r)|^2 + C_0,
\end{equation}
where $\ell$ and $a$ are deduced from the fit to $C_\psi(\Delta r)$, so that $C_0$ is the only free-fit parameter.

The probability density of normalized intensity (Fig. 2a, inset) is also calculated in the frequency domain for the antenna bandwidth and averaged over the acquisitions.
The data were binned (over 40 values in the weak scattering case, and 50 values for strong scattering) from $I/\langle I \rangle = 0$ to $I/\langle I \rangle = \textrm{max}\{I/\langle I \rangle\}$.
The theoretical function used for the fit, which is valid for an incident plane wave, is\cite{Nieuwenhuizen95,vanrossum99}:
\begin{equation}\label{eqS5}
P(I/\langle I \rangle) = \int_0^{\infty}\frac{dv}{v}\int_{-\textrm{i}\infty}^{\textrm{i}\infty}\frac{dx}{2\pi\textrm{i}}\textrm{exp}\left ( \frac{-I/\langle I \rangle}{v} +xv- \Psi(x)) \right ),
\end{equation}
with $\Psi(x) = g\ln^2\left (\sqrt{1+x/g} + \sqrt{x/g} \right )$ and $g$ as a free-fit parameter.

\subsection{Coherent back-scattering}

The angular dependence of the back-scattered field is measured by applying the beamforming method \cite{Aubry2007} to each acquisition.
A controlled phase shift is applied to each receiving channel, which allows the projection of the received signals onto angles along the $x$ or $y$ axes.
To improve the statistics, we take the means over the two angular intensity profiles, so that the results are shown only as a function of one angle $\theta$.F

The dynamic CBS profiles (Fig. 3b, c) are measured for cages C4 and C5, and integrated over time to estimate the static CBS peak (Fig. 3a).
However, our scattering medium is not semi-infinite and the maximum integration time is shorter than the minimum time imposed by the echo from the cage net: $t_{max} = 2d/v_0$ (with $d$ as the depth of the cage).
Nevertheless, our cages are sufficiently large to make the consequent error negligible: a 25\% variation in $t_{max}$ induces 10\% and 5\% variations in the $\ell^*$ estimations for C4 and C5, respectively. 

The theoretical expression used to fit the CBS profile follows from the same diffusion theory as the expression for the mean intensity (\ref{eqS1}):
\begin{equation}\label{eqS6}
\langle I(\theta, t) \rangle = \frac{I_0}{2\pi}\int_{-\infty}^{+\infty}
\frac{z_0}{D}
\left \{\frac{\textrm{e}^{-\gamma_0z'}}{1 + \gamma_0z_0} + \frac{\textrm{e}^{-\gamma z'}}{1 + \gamma z_0}\right \}\textrm{exp}(-\textrm{i}\Omega t)d\Omega,
\end{equation}
where $\gamma^2(\theta,\Omega) = \frac{-\textrm{i}\Omega}{D} + k_0^2\textrm{sin}^2(\theta) + \frac{1}{D\tau_a}$ and $\gamma_0 = \gamma(\theta = 0,\Omega)$.
The fits are first performed for the stationary CBS peak $\langle I(\theta) \rangle$:
\begin{equation}\label{eqS7}
\langle I(\theta) \rangle = 
\int_{0}^{\infty} \langle I(\theta, t) \rangle
dt = 
I_0 \frac{z_0}{D}
\left \{\frac{\textrm{e}^{-\gamma_0 z'}}{1 + \gamma_0z_0} + \frac{\textrm{e}^{-\gamma_1 z'}}{1 + \gamma_1 z_0}\right \},
\end{equation}
where $\gamma_1 = \gamma(\theta,\Omega=0)$. 
$\ell^*$ and $D \tau_a$ are determined from the fit as best-fit parameters.
These parameters are then fixed to fit the dynamic CBS profile $\langle I(\theta, t) \rangle$, with $D$ as a free-fit parameter. These two fits finally yield the energy transport velocity: $v = 3D/\ell^*$.

\subsection{Energy velocity of diffusive waves}

Here we suggest another approach for the calculation of energy velocity based on theories first developed for multiple scattering of light \cite{vanAlbada1991}.
In this problem, we consider the scattering of a plane wave (with wavenumber $k_w=\omega/v_w$) by the object represented on Fig. A1.
As a first approximation, we consider spherical scatterers (the scattering of plane wave by spheroids mainly affects the scattering anisotropy factor and resonant frequencies \cite{Brunet2013}).
The region (0) corresponds to the surrounding sea water, (1) is a thin layer of hard bones and scales, (2) is a thick layer of soft flesh and (3) is the swim bladder (considered as vacuum in this case).

\begin{figure}[h]
%%\vspace*{-0.5cm}
%% \hspace*{-0.55cm}
\includegraphics[width=10cm]{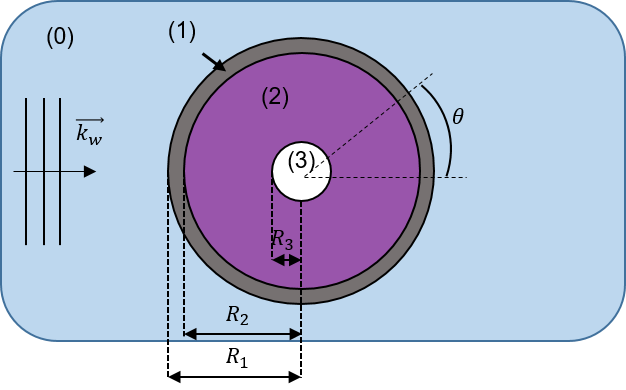}
%%\vspace*{-1.3cm}
\caption{\label{figS2}
Geometry employed to solve the Mie problem and calculate the energy velocity.}
\end{figure}

\begin{figure}[h]
%%\vspace*{-0.5cm}
%% \hspace*{-0.55cm}
\includegraphics[width=15cm]{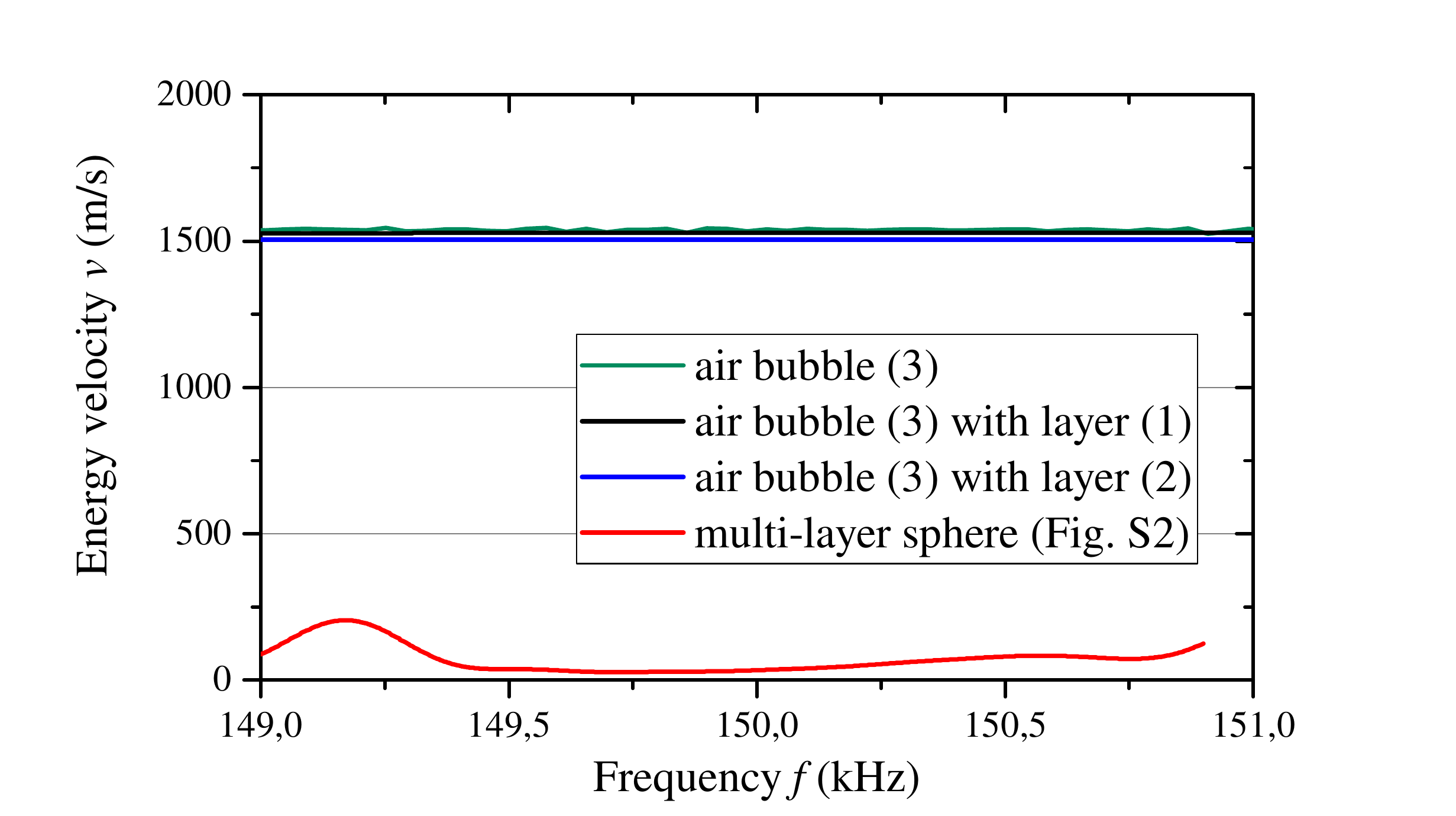}
%%\vspace*{-1.3cm}
\caption{\label{figS3}
Geometry employed to solve the Mie problem and calculate the energy velocity.}
\end{figure}

The scattering amplitude $f(\theta)$ (with $\theta$ the scattering angle) is calculated by solving the Mie problem as detailed in \cite{Hulst1981}.
Thus, the independent scattering approximation (ISA) gives an expression of the effective wavenumber in the fish shoal:
\begin{equation}\label{eqS8}
k^2 = k_w^2 + \frac{16\pi^2}{3}R_1^3\phi f(0)
\end{equation}
The scattering amplitude and effective wavenumber are used for the calculation of energy velocity \cite{vanAlbada1991}:
\begin{equation}\label{eqS9}
v = \frac{v_w^2/v_\textrm{ph}}{1 + \delta},
\end{equation}
with:
\begin{equation}\label{eqS10}
\delta = \frac{8\pi^2}{3}R_1^3\phi\frac{\partial\omega}{\partial k}\left ( \frac{1}{k} \frac{\partial\textrm{Re}[f(0)]}{\partial\omega} \int_0^\pi\textrm{sin}\theta \left | f(\theta) \right |^2 \frac{\partial\varphi(\theta)}{\partial\omega}\textrm{d}\theta \right ),
\end{equation}
$v_\textrm{ph} = \omega/k$ and $f(\theta)=\left | f(\theta) \right |\textrm{e}^{\textrm{j}\varphi(\theta)}$.
Fish size dispersion is taken into account by averaging the  $v$ over different particle sizes with a Gaussian distribution \cite{Mascaro2013}.
A very low size polydispersity ($\sim$ 1 \%) is sufficient to wash out any residual resonant effects that cannot be experimentally observed.
The result of energy velocity calculation (with input parameters detailed on Table \ref{tab2}) is presented on Fig. \ref{figS3} for a small frequency range around the working frequency of our sonar probe.
The resulting energy velocity ($v\sim$ 30 m/s) is found to be very close to both our observation and predictions given by Eq. (4) in the main text.
For comparison, we also plotted on Fig. \ref{figS2} calculations for a simple cloud of air bubble in water and for coated bubbles with materials (1) or (2).
This results show the importance of the multi-layer geometry: the thin layer of bones and scales (1) helps for the wave polarisation conversion (longitudinal-shear waves) while the thick layer of flesh significantly slows down the wave propagation.
The energy equipartition approach employed in the main text assume this polarization conversion and only focus on the flesh shear properties.

\renewcommand{\arraystretch}{1}
\setlength{\tabcolsep}{10pt}
\begin{table}
\begin{center}
\begin{tabular}{|c|c|c|c|c|}
\hline
Region & Radius $R$ (mm) &\begin{tabular}[c]{@{}c@{}} Longitudinal wave \\ speed (m/s) \end{tabular}&\begin{tabular}[c]{@{}c@{}} Shear wave \\ speed (m/s) \end{tabular}& Density (g/cm$^3$)\\ \hline
(0) & \cellcolor[HTML]{9B9B9B} & 1480 & \cellcolor[HTML]{9B9B9B} & $ 1 $ \\ \hline
(1) & 32 & 1600 & 900 & 1.4 \\ \hline
(2) & 30 & 1600 & 10 & 1.1 \\ \hline
(3) & 5 & \cellcolor[HTML]{9B9B9B} & \cellcolor[HTML]{9B9B9B} & 0 \\ \hline
\end{tabular}
\end{center}
\caption{\label{tab2} Input parameters used for the energy velocity calculation with the geometry presented on Fig. \ref{figS3}.}
\end{table}
\renewcommand{\arraystretch}{1}

In conclusion, the very slow diffusive wave transport has been explained simultaneously both by macroscopic approach (by assuming wave energy equipartition) or microscopic description (by calculating scattering properties of an isolated scatterer).
This comparison emphasizes the role of shear wave propagation in the fish body that cannot be considered as a simple air bubble.